\documentclass{emulateapj}
\usepackage{amssymb}
\makeatletter

\newcommand{\Rmnum}[1]{\expandafter\@slowromancap\romannumeral #1@}
\makeatother
\usepackage{amsmath}
\usepackage{array}
\usepackage{bm}
\usepackage{cases}
\usepackage{txfonts}
\usepackage{tipa}
\usepackage{amsfonts}
\usepackage{amssymb}
\usepackage{subfigure}
\usepackage{graphicx}
\usepackage{multirow}
\usepackage{float}
\usepackage{longtable}
\usepackage{graphicx} 
\usepackage{epstopdf}

\begin{document}
\title{Element Abundance Analysis of the Metal-Rich Stellar Halo and High-velocity Thick Disk in the Galaxy }

\author{Haifan Zhu\altaffilmark{1}, Cuihua Du\altaffilmark{2}, Yepeng Yan\altaffilmark{3,1}, Jianrong Shi\altaffilmark{4,2}, Jun Ma\altaffilmark{4,2}, Heidi Jo Newberg\altaffilmark{5}}

\affil{$^{1}$School of Physical Sciences, University of Chinese Academy of Sciences, Beijing 100049, China\\
	$^{2}$College of Astronomy and Space Sciences, University of Chinese Academy of Sciences, Beijing 100049, China; ducuihua@ucas. ac. cn \\          
	$^{3}$Department of Astronomy, Beijing Normal University, Beijing 100875, China\\ 
	$^{4}$Key Laboratory of Optical Astronomy, National Astronomical Observatories, Chinese Academy of Sciences, Beijing 100012, China\\ 
	$^{5}$Department of Physics, Applied Physics and Astronomy, Rensselaer Polytechnic Institute, Troy, NY 12180, USA\\   
}

\begin{abstract}  
Based on the second Gaia data release (DR2) and  APOGEE (DR16) spectroscopic surveys, 
we defined two kinds of star sample: high-velocity thick disk (HVTD) with $v_{\phi}>90 {\rm km/s}$ 
and  metal-rich stellar halo (MRSH) with $v_{\phi}<90 {\rm km/s}$.  
Due to high resolution spectra data from APOGEE (DR16), we can analyze accurately the element abundance distribution of HVTD and MRSH. 
 These elements abundance constituted a multidimensional data space, 
 and we introduced an algorithm method for processing multi-dimensional data to give the result of dimensionality reduction clustering.
 According to chemical properties analysis, we derived that  some HVTD  stars could origin from the thin disk, 
and some MRSH stars from dwarf galaxies,  but those stars which have similar chemical abundance 
characteristics in both sample may form in-situ.

\end{abstract}
\keywords{Galaxy:disk-Galaxy:halo-Galaxy:structure-Galaxy:kinematics-Stars:abundance}

\section{Introduction}
The Halo and thick-disk are the basic components of the Galaxy, and the study 
to their kinematics and chemical abundance could provide important clues
to the Galaxy's formation and evolution history.  
Recent studies \citep{Carollo07,Carollo10,Chen11,Beers12,Kinman12} have shown that the Galactic halo comprises at least two stellar populations: inner-halo and outer halo. 
They have different kinematics, spatial distribution, and chemical composition \citep{Carollo07,Carollo12,Liu18,Conroy19, Bird20}. 
For example, the inner-halo is mainly distributed at distances up to $10-15$ kpc from the Galactic center, 
 and the mean metallicity of the inner-halo range from  [Fe/H] $\sim$ $-1.2$ dex to $-1.7$ dex.
 The outer-halo is mainly distributed at distances up to $15-20$ kpc from the Galactic center, 
 and the mean metallicity of the outer-halo range from  [Fe/H] $\sim$ $-1.9$ dex to $-2.3$ dex \citep[e.g.,][]{Carollo07,An13,An15,Zuo17,Gu15,Gu16,Gu19,Liu18}.
Studies of the detailed chemical abundances and ages of halo
stars have sought to place further constraints on the structure and formation
of the Galactic halo \citep{Naidu20,Sahin20}.  Many studies have shown that there are two chemically distinct 
 stellar populations: an older, high-$\alpha$, and a younger, low-$\alpha$ halo population,
 \citep[e.g.,][]{Nissen10,Nissen11,Bergemann17, Hayes18}. 
 
 Particularly, after the second release of the Gaia (Gaia DR2), 
 a significant number of works have revealed
an even more complex but detailed picture of the Galactic halo.
For example,   \cite{Belokurov18} showed the velocity ellipsoid  becomes strongly anisotropic for the halo stars with $-1.7 <$  [Fe/H]$<-1.0$ dex and local velocity distribution appears highly stretched in the radial direction, taking sausage-like shape, and they suggested that such orbital configurations could show that most of the inner halo stars should be dominated by stars accreted from an ancient massive merger event. This merger event referred as the Gaia-Sausage merger \citep[e.g.,][]{Myeong18,Myeong19,Deason18,Lancaster19}.  
\cite{Helmi18} also demonstrated that the inner halo is dominated by debris from the merger of a dwarf galaxy 10 Gyr ago, and the dwarf galaxy referred as Gaia-Enceladus.
Several other works also found new chemo-dynamical
properties of the stellar halo, that were unknown before Gaia \citep[e.g.,][]{Bonaca17, Bonaca20,
Koppelman19, Belokurov20, Carollo20,Yuan20, Naidu20,Yan20}.
All these observation results imply that accreted stars from satellite galaxies have been suggested to be dominant inner halo component.

The Galactic disks contain a substantial fraction of their baryonic matter and angular momentum, and much of the evolutionary activity. 
 The formation and evolution of disks are therefore very important \citep{Kruit11}. 
 The basic components of the Galactic disk are
the thin-disk and thick-disk populations. The two components differ
not only in their spatial distribution profiles but also in their kinematics, age and metallicity \citep{Kruit11,Ivezi12, Xiang15, Xiang17, Jing16, Peng18, Gandhi19,Han20, Wu21}.
 Stars in the thick disk have the following characteristics compared to stars in the thin disk: 
 older, kinematically hotter, metal-poor and enhanced in $\alpha-$elements \citep[e.g.,][]{Chiba00,Prochaska00,Bensby05}. 
 In addition,  there is evidence for the additional presence of a metal-weak thick-disk (MWTD) population,
rotationally supported, but extending to lower metallicity stars than the canonical thick disk \citep{Morrison90,Beers95,Chiba00,Beers02,Beers14,Carollo19,Yan19}. 
 A key question now is whether these disk components can also be used to account
for the chemical and kinematic measurements for the same stars. 
It turns out that, even with the new data collected, it is not easy to answer it. 
 
Despite the past three decades of thick-disk studies, there is still no consensus on models for
the formation and evolution of thick disk. 
The proposed simulations of thick disk formation can be generally 
divided into four groups: (a) accretion from disrupted satellite galaxies \citep{Ab03},
 (b) heating of the pre-existing thin disk due to minor mergers \citep{Quinn93}, (c) in-situ triggered star formation during 
and after a gas-rich merger \citep{Brook04,Sales09}, (d) in-situ formation through radial migration \citep{SB02, SB09a, SB09b,Sch11}.

These currently discussed models of
formation mechanisms for the thick disk predict various trends between
the kinematics properties and the metallicity of disk stars.
For example, some simulations of accretion from disrupted satellite galaxies can help to explain why there are so many old stars in circular orbits in the outskirts of galaxies 
and why the specific angular momentum and radial extent of the thick disk and thin disk are comparable although their ages are significantly 
different \citep{Ab03}.  Furthermore, it provides an explanation for the dynamical and evolutionary distinction  between the thick and thin disk components:  the thick disk is mostly tidal debris from disrupted satellites, while the young thin disk consists mostly of stars formed
in situ after the merging activity abates \citep{Ab03}.
Some simulations of the thick disk formation via the accretion of 
satellites onto a pre-existing thin disk, showed that mergers with $10\%-20\%$ mass of the host lead to the the formation of thick disk \citep{Villalobos08}.
Using a different numerical implementation of the radial migration scenario,  Sch{\"o}enrich, and Binney \citep{SB09a, SB09b, Sch11} showed 
radial migration plays an important role in the local Solar Neighborhood.  \cite{Feuillet2019} studied the spatial changes in the [$\alpha$/M]-age relation and [M/H]-age relations of the Milky Way disk. The results are important constraints to Galactic simulations and chemical evolution models. 

\cite{Kordopatis20} used the chemo-dynamical method to give conclusions related to the formation of thick disks.
They think the key to understanding the effect of the
past accretions on the properties of the thick disk is the super-solar
metallicity counter-rotating population. This population will provide us with a reliable sample of locally born retrograde 
stars in order to determine the exact time and weigh this merger \citep{Grand20}.

Previous studies showed that most of halo stars are metal-poor.  With the release of more survey data,
 some recent works have revealed a large number of metal-rich halo stars ([Fe/H]$>-1$ dex) \citep[e.g.,][]{Nissen10,Nissen11,Bonaca17,Posti18,Yan20}.  
\cite{Bonaca17} use the first Gaia data release, the Radial Velocity Experiment (RAVE, \citealt{stei06}) and \textbf{the Apache Point Observatory Galactic Evolution Experiment (APOGEE, \citealt{Eisenstein2011})}  to find that half of their halo sample is comprised of stars with [Fe/H]$>-1$ dex, 
and they proposed that metal-rich halo stars in the solar neighborhood actually formed in situ within the Galactic disk. It is possible that these stars have undergone radial migration that caused changes in their orbits.  \cite{Yan20} use the second Gaia data release (DR2), 
combined with the ongoing Large Sky Area Multi-Object Fiber Spectroscopic
Telescope survey (LAMOST, also called Guoshoujing Telescope, \citealt{zhao12}) and APOGEE to show that there 
exist a high-velocity thick disk (HVTD) with $v_{\phi}>90{\rm km/s}$ and a metal-rich stellar halo (MRSH) with $v_{\phi}<90{\rm km/s}$ in the Galaxy.  
But the details of how high-velocity thick disk (HVTD) and metal-rich stellar halo (MRSH) form are not yet understood.

In this work, we use the APOGEE DR16 and Gaia DR2 data to obtain samples of HVTD and MRSH to study the detailed abundance characteristics and give their possible origins.
 This paper is organized as follows:  Sect.\ref{section2} introduces the data selection and the method of distance determination, and give the division of HVTD and MRSH.
 Section \ref{section3} introduces the dimensionality reduction algorithm 
 to present the element abundances analysis. 
  In Sect.\ref{section4}, we analyze the results and compare them with some previous works, and give the possible origin of HVTD and MRSH. 
  The summary and conclusions are given in Sect.\ref{section5}.

\section{Data}
\label{section2}
\subsection{Data selection}
The Apache Point Observatory Galactic Evolution
Experiment (APOGEE), part of the Sloan Digital Sky Survey \Rmnum{3},
is a near-infrared (H-band; 1.51-1.70 ${\rm \mu m}$) and high-resolution (R $\sim$ 22,500) spectroscopic survey \citep{Zasowski13}. 
In this study, we use the
data from the sixteenth SDSS Data Release (DR16) and  it provides accurate radial velocities, stellar parameters, and
chemical abundances of about 26 chemical species for
430 000 stars covering both the northern and southern sky \citep{Jonsson20}.

The second Gaia data release, Gaia DR2, provides high-precision positions, parallaxes, and proper motions for 1.3 billion sources brighter than magnitude $G \sim 21$ mag. More detailed information about Gaia can be found in \cite{Gaia18a, Gaia18b}.  
In this study,  we get the high-resolution sample stars by 
cross-matching between the APOGEE DR16 and Gaia DR2 catalog. Stellar 
parameters (metallicity abundances, radial velocity, effective temperature, and surface gravity)
are from the APOGEE DR16. Other necessary parameters (position, proper motion, and parallax) are from
the Gaia DR2 catalog.  

In order to obtain reliable results, we utilize the following selection criteria:
\begin{itemize}
	\item parallax uncertainties $< 20$\%
	\item proper motion error $<0. 2$ mas/year
	\item radial velocity uncertainties $<10$  $\rm km/s$
	\item $ \text{[Fe/H]}$ error $<0. 2$ dex
	\item S/N $> 20$ in the G-band
\end{itemize}
\subsection{Distance and Velocity determination }

\cite{Bailer-Jones15} discussed that the inversion of the parallax to obtain distance is not appropriate when the relative parallax error is more than 20 percent. 
We use the Bayesian approach \citep{Bailer-Jones15, AB16a,AB16b, luri18,Yan19} to derive stellar distance.

 According to Bayesian formula,  the posterior probability can be written as follows:
 \begin{equation}
	\begin{split}
	P(\bm{{\rm \theta}}|\bm{{\rm x}})\propto & P(\bm{{\rm x}}|\bm{{\rm \theta}})P(\bm{{\rm \theta}})
	\\=& \textrm{exp}[-\frac{1}{2}(\bm{{\rm x}}-\bm{{\rm m}}(\bm{{\rm \theta}}))^{\rm T}\Sigma^{-1}(\bm{{\rm x}}-\bm{{\rm m}}(\bm{{\rm \theta}}))]P(\bm{{\rm \theta}}),
\end{split}
\label{eq1}
\end{equation}
where $P(\bm{{\rm x}}|\bm{{\rm \theta}})$ is the likelihood, the symbol $\bm{\theta}$ represent 
the parameter vectors which consists of heliocentric distance
($d$), tangential speed ($v$), and travel direction ($\phi$, increasing
anti-clockwise from north), written as
\begin{equation}
	\bm{{\rm \theta}} = (d, v, \phi)^{\rm T}.
\end{equation}
$\rm x$ is the observed data vector which consists of the parallax ($\overline{\omega}$), proper motion in R.A. ($\mu_{\alpha^*}$) and DEC.
($\mu_{\delta}$), written as 
\begin{equation}
	\bm{{\rm x}} = (\varpi, \mu_{\alpha^*}, \mu_{\delta})^{\rm T}.
\end{equation}
The likelihood probability is a multidimensional Gaussian distribution
centered on $\bm{m}$ as we can see in the formula \eqref{eq1}.
$\Sigma$ is covariance matrix
\begin{equation}
\begin{split}
&{\rm \Sigma}=\\
&\begin{pmatrix} \sigma^2_{\varpi} & \sigma_{\varpi}\sigma_{ \mu_{\alpha^*}}\rho(\varpi,\mu_{\alpha^*}) & \sigma_{\varpi}\sigma_{\mu_{\delta}}\rho(\varpi,\mu_{\delta}) 
	\\\sigma_{\varpi}\sigma_{ \mu_{\alpha^*}}\rho(\varpi,\mu_{\alpha^*}) & \sigma^2_{\mu_{\alpha^*}} & \sigma_{ \mu_{\alpha^*}}\sigma_{\mu_{\delta}}\rho(\mu_{\alpha^*},\mu_{\delta}) 
	 \\ \sigma_{\varpi}\sigma_{\mu_{\delta}}\rho(\varpi,\mu_{\delta}) & \sigma_{ \mu_{\alpha^*}}\sigma_{\mu_{\delta}}\rho(\mu_{\alpha^*},\mu_{\delta}) & \sigma_{\mu_{\delta}}^2  \end{pmatrix}.
\end{split}
\end{equation}
where $\rho(i,j)$ denotes the correlation coefficient between $i$ and $j$
and $\sigma_{k}$ denotes the standard deviation of parameters $k$.
$\rm{{\bm m}}$ represents a set of theoretical values predicted by our model, written as
\begin{equation}
\bm{{\rm m}}= (\frac{10^3}{d}, {c} \frac{10^3v  \sin\phi}{d}, {c} \frac{10^3v  \cos\phi}{d})^{\rm T}
\end{equation}
where $c = ({\rm pc\cdot mas\cdot yr^{-1} })/({\rm 4.74\cdot km\ s^{-1}})$.

$P(\bm{{\rm \theta}})$ is the prior distribution given by \cite{luri18}
\begin{equation}
	P(\bm{{\rm \theta}})=P(d)P(v)P(\phi),	
\end{equation}
with
\begin{align}
	P(d) &\propto \begin{cases} d^2e^{-d/L(a,b)}& d>0\\ 0& d\le 0 \end{cases}\\
	P(v) &\propto \begin{cases} (\frac{v}{v_{\rm max}})^{\alpha-1}(1-\frac{v}{v_{\rm max}})^{\beta-1}& {\rm if}\ 0 \le v \le v_{\rm max}\\ 0& \text{otherwise} \end{cases}\\
	P(\phi) &\propto \frac{1}{2\pi}
\end{align}
$L(a,b)$ is length scale of Galactic longitude- and
latitude-dependent \citep{Bailer-Jones15}. Here we take $\alpha=2,~ \beta=3, ~v_{max}=750~ \rm km/s$.
Then we get the posterior distribution through the Markov chain
Monte Carlo (MCMC) sampler EMCEE \citep{for13}. We run each
chain using 100 walkers and 100 steps, for a total of 10,000
random samples drawn from the posterior distribution. Finally, we adopt the data to calculate the 
kinematic parameters by using Astropy \citep{as13,as18} and galpy \citep{Bovy15}.
Astropy provides many functions to calculate the parameters and 
coordinate transformations. 
Galpy is a Python package for galactic dynamics and it  supports orbit integration in various potentials.

When calculating with galpy, we choose the potential model from MWPotential2014 \citep{Bovy15}, 
the distance from the Sun to the 
Galactic center is  $\rm 8. 2$ kpc and height above the plane is about 15 pc \citep{Bland-Hawthorn16}. 
When deriving velocity, we adopt $\rm v_{LSR} = 232. 8 ~\rm km ~s^{-1}$	\citep{McMillan17}. 
The velocity of the Sun with respect to the the local standard of rest (LSR) are
$(U_{\odot}, V_{\odot}, W_{\odot}) = (10., 11., 7.)$ km s$^{-1}$ \citep{Bland-Hawthorn16}.
Using the above, we can get the coordinate position, speed and orbit parameters.

\subsection{Determination of HVTD and MRSH}

\begin{figure}[]
	\centering
	\includegraphics[width=0.5\textwidth]{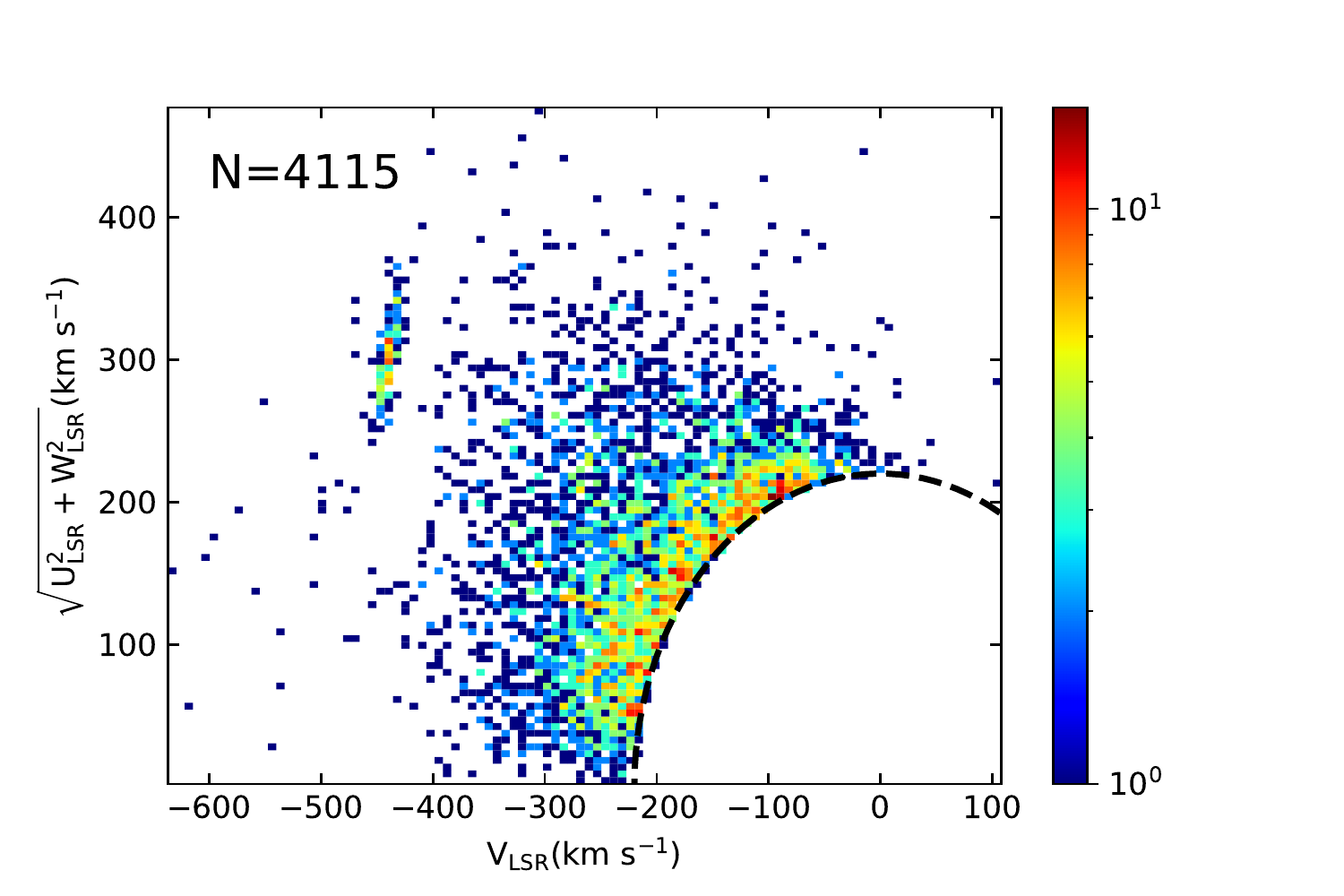}
	\caption{ Toomre diagram of our high-velocity sample stars. 	
	  The black dashed line represents the 
	 total spatial velocity  
	 $v_{{\rm tot}} =220$ ${\rm km\ s^{-1}}$( $\rm v_{ LSR}=232. 8\ {\rm km\ s^{-1}}$). 
	 Our high-velocity sample stars are defined as $v_{{\rm tot}}>220$ ${\rm km\ s^{-1}}$. 
	 The color bar represents the number of stars. }
	\label{figure1}
\end{figure}

\begin{figure}[]
	\centering
	\includegraphics[width=0.5\textwidth]{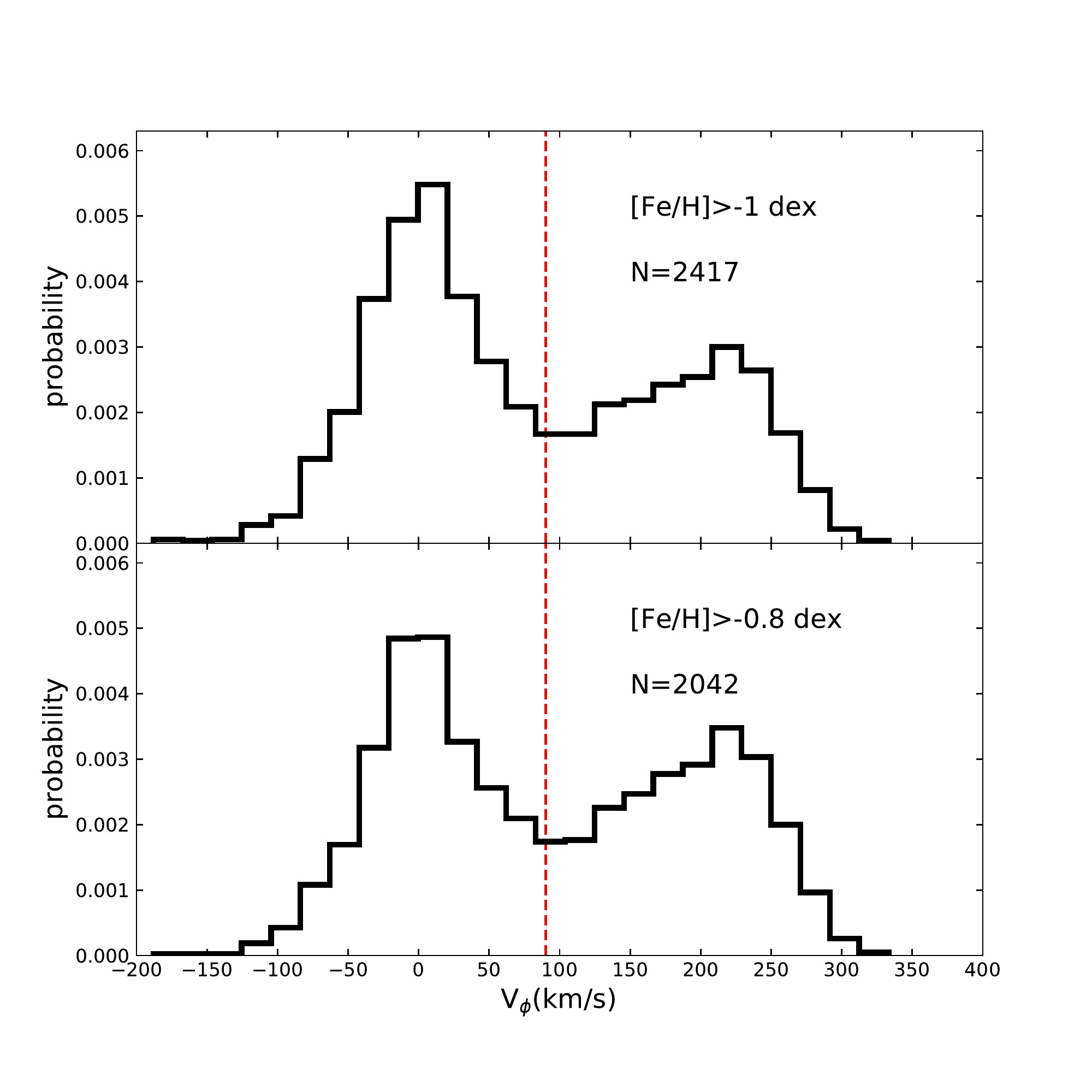}
	\caption{ The rotational velocity distribution of high-velocity sample stars.
	 Red dashed line is the division between HVTD and MRSH ($V_{\phi}=90 ~\rm km~s^{-1}$). 
	  }
	\label{figure2}
\end{figure}

\begin{figure}[]
	\centering
	
	\includegraphics[width=0.5\textwidth]{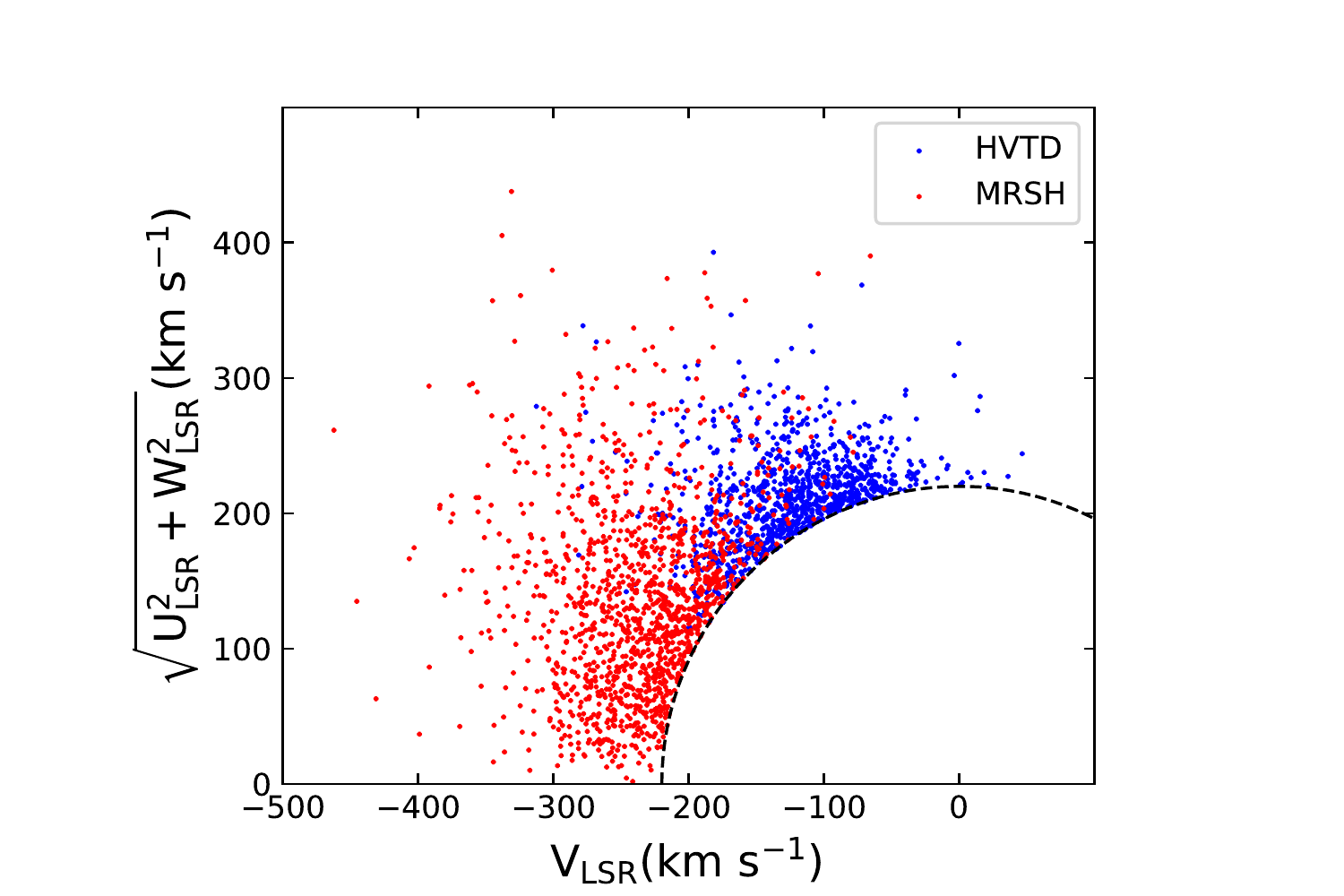}
	\caption{Toomre diagram of the HVTD (blue dots) sample and the MRSH (red dots) sample. 
	The black dashed line represents the total spatial velocity  
	$v_{{\rm tot}} =220$ ${\rm km\ s^{-1}}$.
	The two samples have a clear boundary and there are some overlapping parts at the
	 boundary.}
	\label{figure3}
\end{figure}

In this study, we first get 4115 high velocity sample stars with $\rm v_{tot}>220~ \rm km~s^{-1}$. 
The sample stars distribution in the Toomre diagram is presented in Figure~\ref{figure1}. 
In order to determine the  HVTD component and the MRSH component,
 we also give the rotation velocity distribution of the sample stars with different metallicity selection in Figure~\ref{figure2}.
  According to \cite{Yan20}, two-dimensional Gaussian fitting  implied the existence of the HVTD and MRSH in the high-velocity sample, 
  and they gave a definition of the division of the two 
 components: HVTD with $v_{\phi}>90 ~\rm km~s^{-1}$ 
 and [Fe/H]$>-1$ dex, MRSH with $v_{\phi}<90 ~\rm km~s^{-1}$ 
 and [Fe/H]$>-1$ dex.  Figure~\ref{figure3} gives the Toomre diagram. We can see that the distribution of HVTD and MRSH samples is quite 
different in the Toomre diagram. There is a small overlap near the dividing line. 

The standard for division of two samples is derived from the characteristics of the bimodal distribution of the overall sample. 
It is not easy to obtain a clear boundary for the sample distribution.  In this work, we directly adopted the previous division method. 

The rotational velocity from $75 ~\rm km~s^{-1}$ to $125 ~\rm km~s^{-1}$ between the two peaks of the velocity distribution is suitable as division in Figure~\ref{figure2}.   Different dividing lines will affect the number of samples of MRSH and HVTD, but there are only 204 sample stars in this range for [Fe/H]$>-1$ dex and 176  sample stars for [Fe/H]$>-0.8$ dex. 
Furthermore, there are  65 stars in interval of $v_{\phi}$  [$75 ~\rm km~s^{-1}$, $90 ~\rm km~s^{-1}$],
84  stars in interval [$90 ~\rm km~s^{-1}$, $110 ~\rm km~s^{-1}$],
55 stars in interval [$110 ~\rm km~s^{-1}$, $125 ~\rm km~s^{-1}$] with [Fe/H]$>-1$ dex. For [Fe/H]$>-0.8$ dex, the sample is even smaller. So even if changing the classification standard in this range, only classification of a very small sample of stars will be changed, 
and these changes will not have a substantial effect on the subsequent study. 

\subsection{Metallicity Distribution and Gradient}

\begin{figure}[]
	\centering
	\includegraphics[width=0.5\textwidth]{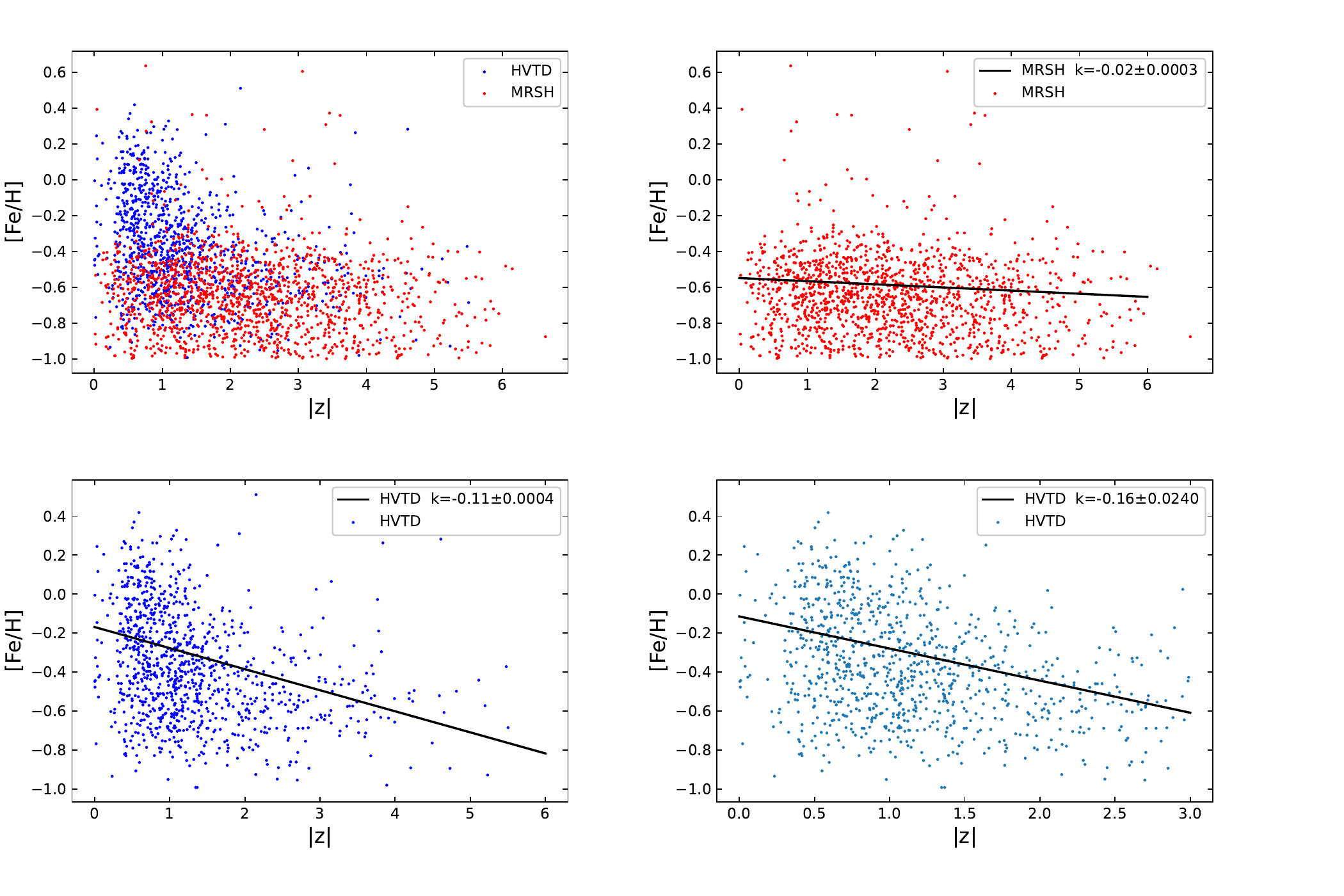}

	\caption{Top left panel: Metallicity distribution as 
	function of $\rm |z|$ (kpc) for HVTD marked by the blue dots and MRSH marked by the red dots. 
	The distribution of HVTD is relatively concentrated, and the 
	distribution of MRSH is relatively uniform. 
	Top right panel: Metallicity distribution as 
	function of $\rm |z|$ of MRSH, $d[{\rm Fe/H}]/d|z| =-0.02 \pm 0. 0003 ~{\rm dex /kpc}$. 
	Bottom left panel: Metallicity distribution as 
	function of $\rm |z|$ for the HVTD, $d[{\rm Fe/H}]/d|z| =-0.11 \pm 0. 0004 ~{\rm dex /kpc}$.
	Bottom right panel: Metallicity distribution as 
	function of $\rm |z|$ for the HVTD in the range $|z|<3 ~{\rm kpc}$,
	$d[{\rm Fe/H}]/d|z| =-0. 16 \pm 0. 0008~{\rm dex/kpc}$.
	}
	\label{figure4}
\end{figure}

According to previous research, we use kinematic information to divide the sample stars into HVTD and MRSH.
Figure~\ref{figure4} shows the metallicity distribution in 
the vertical distance \textbf{$|z|$} of the two samples. 
 We can notice that the distribution of HVTD is 
 relatively concentrated, mainly in the range $<3$ kpc, and the 
distribution of MRSH is relatively uniform. 

We use maximum likelihood estimation to derive the metallicity gradient in the $|z|$ direction.  
For a given sample with total number of stars $N$, the data points conform to the linear form $ \text{[Fe/H]}_{i}= kz_{i} + b$. 
 In this model, if position $z_{i}$, \textbf{$\text{[Fe/H]}_{i}$} uncertainty $\sigma_{i}$ ,  slope $k$ (also $d[\rm Fe/H]$$/dz$), and intercept $b$ are given, 
the distribution $P(\text{[Fe/H]}_{i}| z_{i},\sigma_{i}, k, b)$ for $\text{[Fe/H]}_{i}$ is
\begin{equation}
	P(\text{[Fe/H]}_{i}| z_{i},\sigma_{i},k, b)=\frac{1}{\sqrt{2\pi\sigma_{i}}}exp(-\frac{(\text{[Fe/H]}_{i}-kz_{i}-b)^2}{2\sigma_{i}^2})
\end{equation}
the likelihood $L$ is 
\begin{equation}
	L=\prod_{i=1}^{N}P(\text{[Fe/H]}_{i}| z_{i},\sigma_{i},k, b).
\end{equation}

We aim to derive the parameter value ($k, b$) that maximizes
the likelihood function.  Here, we use the Bayesian approach 
to derive the uncertainty of the metallicity vertical gradient.
The metallicity vertical gradient of the HVTD is
 $d\text{[Fe/H]}/d|z|=-0. 11\pm 0. 0004 ~{\rm dex/kpc}$ 
 as shown in  
the Figure~\ref{figure4} bottom left panel.
The HVTD is mainly distributed in
 $z<3$ kpc,  we calculated the metallicity 
vertical gradient in this range, 
$d\text{[Fe/H]}/d|z|=-0. 16 \pm 0. 0008~{\rm dex/kpc}$. 

This result is roughly consisted with some previous result of the thick disk.  For example, \cite{Chen11}
used RHB stars with range in $0. 5-3 ~{\rm kpc}$
to give the gradient for the thick disk   $d\text{[Fe/H]}/d|z|=-0. 12\pm 0. 01~{\rm dex/kpc}$. 
\cite{Bilir12}  used red clump stars to derive that the vertical metallicity gradient for the thick disk is close
to zero. \cite{Mikolaitis14} given $d\text{[Fe/H]}/d|z|=-0. 072\pm0. 006~{\rm dex/kpc}$.
\cite{Li17} given the result $d\text{[Fe/H]}/d|z|=-0. 164\pm0. 010~{\rm dex/kpc}$.
\cite{Tun19}  reported $d\text{[Fe/H]}/d|z|=-0. 164\pm0. 014~{\rm dex/kpc}$ with range $6<R<10~{\rm dex/kpc}$  and $2<|z|< 5 ~{\rm dex/kpc}$.
\cite{Yan19} reported $d\text{[Fe/H]}/d|z|=-0. 074\pm0. 0009~{\rm dex/kpc}$ for the thick disk. 
According to these studies, the gradient of our HVTD sample is 
very similar to the thick disk, which implies the HVTD sample have 
very obvious characteristics of the thick disk.  

In the top right plane of Figure~\ref{figure4}, 
the metallicity vertical gradient of the MRSH is $d\text{[Fe/H]}/d|z|=-0. 02\pm 0. 0003~{\rm dex/kpc}$,  which is a very flat gradient. 
This result is roughly consisted with previous results of halo, \cite{Tun19} given the vertical metallicity gradients
$d\text{[Fe/H]}/d|z|=-0. 023\pm0. 006~{\rm dex/kpc}$ in the interval $6<R<10$ kpc. \cite{Peng12, Peng13} derived that the vertical gradient 
$d\text{[Fe/H]}/d|z|=-0. 05\pm 0. 04~{\rm dex/kpc}$ and $d\text{[Fe/H]}/d|z|=-0. 03\pm 0. 02~{\rm dex/kpc}$  for distance $5 < z \leq 14$ kpc and they concluded that there is little or no gradient in the halo. Other studies also given the gradient closed to zero at larger galactocentric distances for the halo.

\section{The Chemical Elements  Abundance Analysis}
\label{section3}
\subsection{Stellar Elements Abundance distribution}

\begin{figure*}[]
	\centering
	\includegraphics[width=1\textwidth]{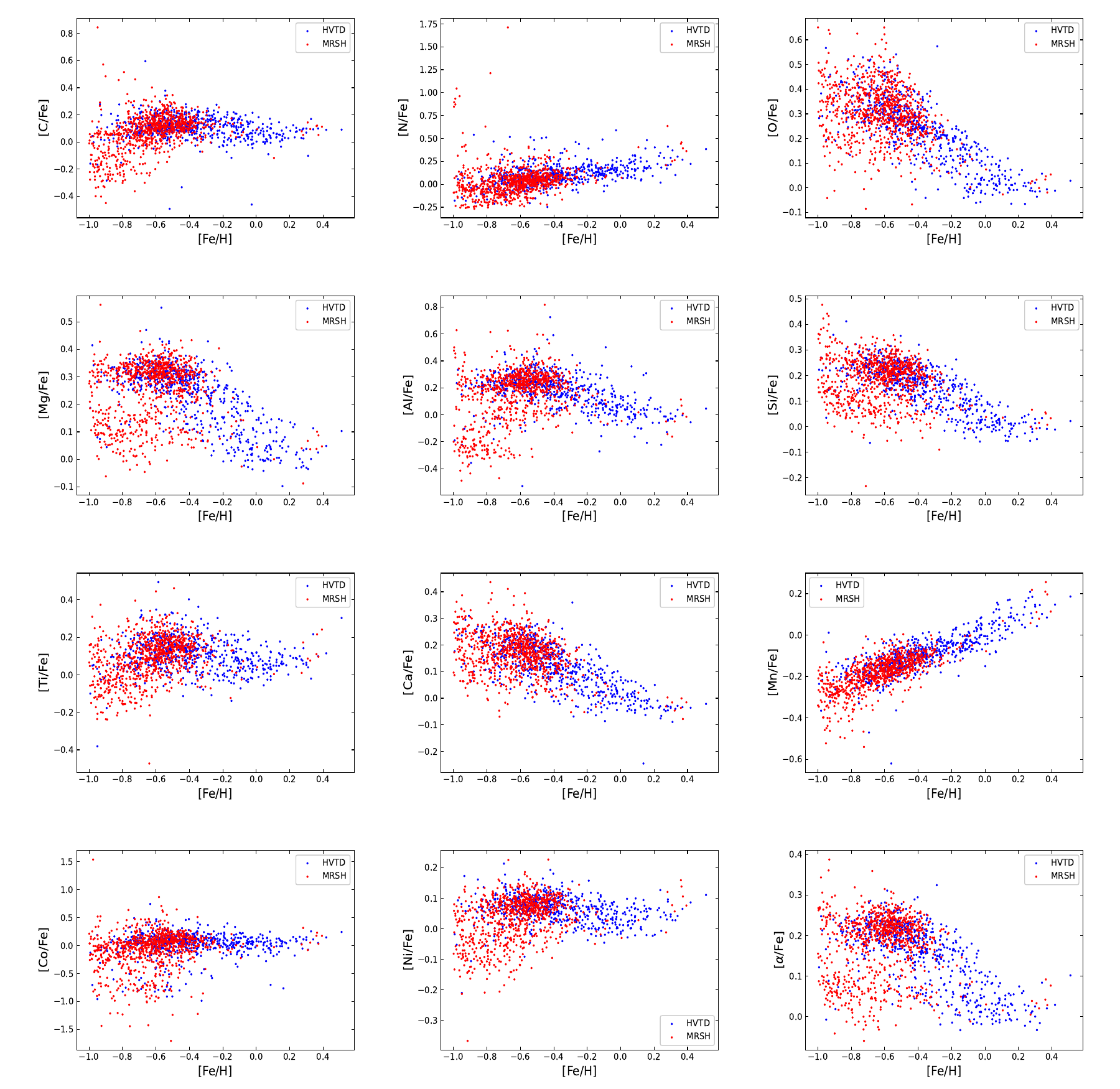}

	\caption{Element abundance distribution versus [Fe/H]. 
	We removed some obvious inaccurate data. 
	The high-velocity thick disk (HVTD, marked by the blue dots) 
	and metal-rich stellar halo (MRSH, marked by the red dots). }
	\label{figure5}
\end{figure*}

APOGEE DR16 provides a large number of chemical elements abundance 
of stars. \cite{Jonsson20}  points out  some parameters 
are not accurate enough and should be selected according to the effective temperature. We selected 11 elements and $\alpha-$element as a function of $\rm [Fe/H]$ 
to study the chemical properties and origins of sample stars. 
These elements abundance distributions are given in Figure~\ref{figure5}. We also derived the important 
parameter  $\rm[\alpha/Fe]$, where $\alpha$ refers to the average 
abundance of Mg, Si, Ca, Ti. The $\alpha-$elements are often used as 
an indicator of time scale in studying the history of star formation. 
This is mainly due to the $\alpha-$elements produced in type II 
supernovae explosion on a relatively short time scale ($10^7$ years).
While iron is produced on a much longer time scale ($10^9$ years) by 
type Ia supernovae explosion. So the analysis of the $\alpha-$elements 
is particularly important.

C, N and O abundance are also important for probing the star formation
 history.  For example, \cite{Masseron15} used the variations of 
 carbon and nitrogen abundances of stars in the thin disk and thick disk of the Galaxy to gather information on the relative ages of stars.  
 \cite{Martig16} used carbon and nitrogen to infer the mass of the stars. Although the derived ages 
are with rms errors of 40 percent, these element abundance is still important to study the formation history.

As shown in each panel of Figure~\ref{figure5}, there are 
a large number of overlapping areas in these distributions. 
In areas with rich metallicity, most sample stars belong to the HVTD. 
The MRSH sample stars are mainly distributed in the range with ${\rm [Fe/H]}<-0.2$ dex.
In addition, from the distribution in the $\rm [X/Fe]$  vs. [Fe/H] (X refers the selected elements), 
we noticed that some chemical elements display centralized distribution, like $\rm N, Ca, Si, Mn$ for the HVTD and MRSH, 
as well $\rm C, Al, Co$ for the HVTD, but some elements appear scattered distribution, like $\rm O, Mg, Ti,Ni$ for the HVTD and MRSH.

\subsection{Data Dimensionality Reduction}

We selected 11 elements (C, N, O, Mg, Al, Si, Ca, Ti, Mn, Co, Ni) which are accurate in APOGEE DR16. 
These elements with [Fe/H] constitute a high-dimensional data space,
and comprehensive understanding and interpreting  of multi-dimensional abundance 
distribution is not easy.  We try to reduce 
dimensionality of multi-dimensional data.  The common data dimensionality
 reduction method is principal component analysis (PCA) algorithm, which is a linear 
algorithm. But the results of this algorithm are not reliable when 
the data has complex connections. So for highly-correlated datasets
 (like chemical-abundance distribution in this work), non-linear dimensionality reduction processing algorithm is necessary.

Here, we use t-distributed stochastic neighbor 
embedding method (t-SNE;\citealt{van08}) which is a non-linear machine-learning algorithm, and it can reduce N-dimensional data 
to a 2D plane. In this work, each star has 11 elements  abundances and metallicity [Fe/H] . 
These elements constitute a high-dimensional data space (Fe, C, N, O, Mg, Al, Si, Ca, Ti, Mn, Co, Ni), 
each element is a dimension. The t-SNE algorithm can reduce each sample star with 12-dimensional data 
characteristics to two dimensions and represent it with a set of coordinates, 
and use a set of two-dimensional plane coordinates to make the results easier to visualize. 
Data points (stars) with similar characteristics on this 
plane will gather together. 

The principle of the algorithm is: given a $N$ high-dimensional data points $\mathbf{x_1},\cdots,\mathbf{x}_{N}$ (here represent abundances), 
t-SNE first computes probabilities $p_{ij}$ that are proportional to the similarity of objects $\mathbf{x}_{i}$ and $\mathbf{x}_{j}$, as follows:
\begin{equation}
	p_{j|i}=\frac{exp(-||\mathbf{x}_{i}-\mathbf{x}_{j}||^2/2\sigma_{i}^2)}{\sum_{i\neq k} exp(-||\mathbf{x}_{i}-\mathbf{x}_{k}||^2/2\sigma_{i}^2)},
\end{equation}
and define symmetrized similarity :
\begin{equation}
	p_{ij}=\frac{p_{j|i}+p_{i|j}}{2N}.
\end{equation}
t-SNE aims to learn a d-dimensional map $\mathbf {y} _{1},\dots ,\mathbf{y}_{N}$ (with $\mathbf {y} _{i}\in \mathbb {R} ^{d}$) that reflects 
the similarities $p_{ij}$ as well as possible.  

For the low-dimensional counterparts $\mathbf y_{i}$ and $\mathbf y_{j}$ of the
high-dimensional data points $\mathbf x_{i}$ and $\mathbf x_{j}$, it is 
possible to compute a similar conditional probability,
which is denoted by $q_{j|i}$.  It reflects the similarities  between two points  $\mathbf y_{i}$ and $\mathbf y_{j}$ in the low-dimensional map.
For the low-dimensional, we choose the variance of the Gaussian 
to $1/\sqrt{2}$
\begin{equation}
	q_{j|i}=\frac{exp(-||\mathbf{y}_{i}-\mathbf{y}_{j}||^2)}{\sum_{i\neq k} exp(-||\mathbf{y}_{i}-\mathbf{y}_{k}||^2)}.
\end{equation}
In the case of symmetry,
\begin{equation}
	q_{ij}=\frac{exp(-||\mathbf{y}_{i}-\mathbf{y}_{j}||^2)}{\sum_{k\neq l} exp(-||\mathbf{y}_{k}-\mathbf{y}_{l}||^2)}.
\end{equation}
In t-SNE algorithm, we employ a Student t-distribution with one degree of freedom 
as the heavy-tailed distribution in the low-dimensional map. 
Using this distribution, the joint probabilities $q_{i|j}$ are defined as
\begin{equation}
	q_{ij}=\frac{(1+||\mathbf{y}_{i}-\mathbf{y}_{j}||^2)^{-1}}{\sum_{k\neq l} (1+||\mathbf{y}_{k}-\mathbf{y}_{l}||^2)^{-1}}.
\end{equation}

The locations of the points $\mathbf {y} _{i}$ in the map are determined by
 minimizing the (non-symmetric) Kullback-Leibler divergence of the distribution 
 $P$ from the distribution $Q$, that is:
\begin{equation}
	KL(P||Q)=\sum_{i\neq j}p_{ij}log\frac{p_{ij}}{q_{ij}}.
\end{equation}
The minimization of the Kullback-Leibler divergence is performed by using gradient descent. 
The result of this optimization
is a 2D plane that reflects the similarities between the
high-dimensional data points \citep{van08}.

We use scikit-learn \citep{Pedregosa11} packages in python to implement this algorithm.
The method has one main parameter perplexity, $p$, which governs the
 bandwidth of the Gaussian kernels $\sigma_{i}$
and appears in the similarities $p_{ij}$, in this work we take $p=50$.
Another parameters is learning rate and 
set to 1, and the early-exaggeration 
parameter is not important, so we take the default value, and  random state in scikit-learn  packages is set to 1.

\begin{figure}[]
	\subfigure{\includegraphics[width=1\hsize]{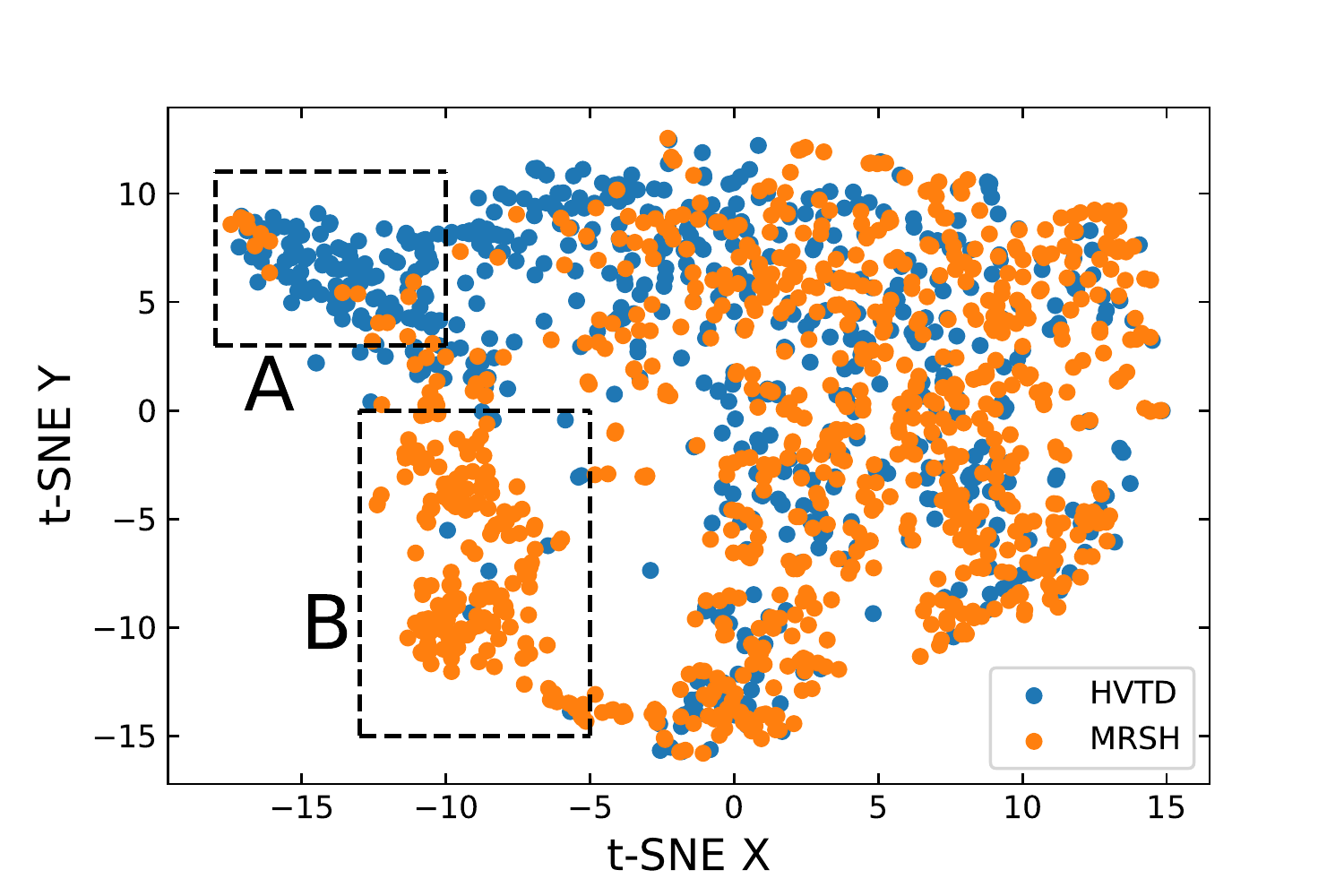}}
	\caption{Dimensionality reduction result,  the HVTD sample is marked by blue dots and the MRSH sample is marked by orange dots. The result contains two parts with a single component (HVTD and MRSH), marked A and B, respectively. }
	\label{figure6}
\end{figure}

\begin{figure*}[]

	\includegraphics[width=1\textwidth]{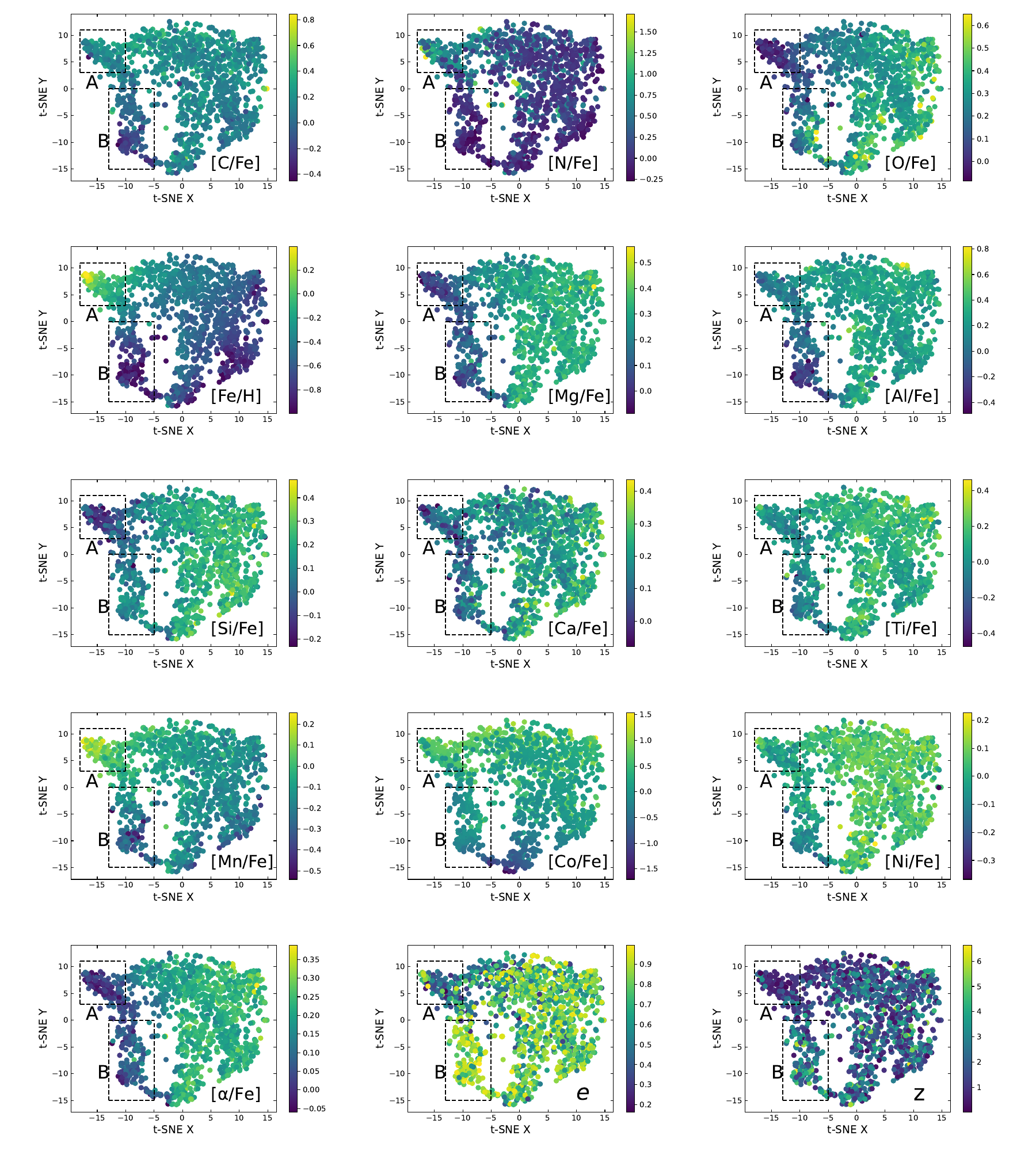}

		\caption{t-SNE result, color-coded by chemical abundance, eccentricity $\rm{e}$ and vertical distance $\rm{|z|}$. 
		We only used [Fe/H] and [X/Fe] ratios to get t-SNE
		result.}
		\label{figure7}
	\end{figure*}

We give the final result of this algorithm in Figure~\ref{figure6}, the HVTD sample is marked by blue dots and the MRSH sample is marked by orange dots. 
At the same time, we have drawn the distribution of different parameters ($\rm[X/Fe]$, $\rm{[Fe/H]}$, eccentricity $\rm{e}$ and vertical distance $\rm{|z|}$) in Figure~\ref{figure7}.

	The entire sample is divided into three areas in Figure~\ref{figure6} :
	\begin{itemize}
		\item Area A in HVTD
		\item Area B in MRSH
		\item Overlapping area of HVTD and MRSH
	\end{itemize}
	Area A and B denote two independent parts in the dimensionality reduction result, which mainly contains HVTD and MRSH, respectively.

\section{Discussion about the  ORIGINS OF THE MRSH AND HVTD}
\label{section4}

Some previous works suggested some possible origin about those metal-rich halo stars \citep[e.g.,][]{Bonaca17,gall19,Belokurov19,Yan20}. 
For example, \cite{Bonaca17} proposed that metal-rich halo stars within $3 \rm kpc$ 
 from the sun may have formed in situ, rather than having been accreted from satellite systems, 
 and these metal-rich halo stars have likely undergone substantial radial
migration or heating. \cite{Yan20} proposed that for the young stars ($<$ 9 Gyr), their formation may not
be affected by the Gaia-Sausage merger. The MRSH stars were likely born in-situ rather than accreted from the Gaia-Sausage merger. 
But for the old stars formed in-situ ($>$ 9 Gyr), the Gaia-Sausage
merger event may have a major effect on their formation. The
MRSH stars may form in an old proto-disk, possibly dynamically heated by the Gaia-Sausage merger, and subsequently
be kicked out to the halo.

In this study, based on the dimensionality reduction results,  we found there are obvious regional chemical characteristics, as shown in Figure~\ref{figure6}. 
The stars in zone B are mainly MRSH, the stellar metallicity in this 
region lies  between -1 dex and -0.6 dex.  But it still has lower $\alpha$-element 
abundances. The $\rm [Mg/Fe]$, $\rm [Al/Fe]$, 
[$\alpha$/Fe] here is obviously much smaller than the overlap area.
It is very similar to the chemical characteristics of dwarf galaxies.
 \cite{Letarte10} showed such element abundance characteristics for field stars of dwarf galaxies (Fornax).
\cite{Hawkins15} also showed in more details the abundance patterns of the elements as a function of metallicity  
for the disks stars, halo stars and dwarf galaxy (Fornax), which clearly implied lower $\alpha$-element with [Fe/H] from -1.0 dex to -0.5 dex for dwarf galaxies.
\cite{Fattahi19} revealed the origin of the metal-rich halo 
stars ($\rm [Fe/H] \sim -1 ~{\rm dex}$) with highly eccentric orbits (high orbital anisotropy, $\beta > 0.8$), by tracing 
their stars back to the epoch of accretion and showed these stars could come from 
a single dwarf galaxy.  Those sample stars in area B also have extremely high 
eccentricities in the last row of the middle column of Figure~\ref{figure7} .  
Therefore, we conclude that those stars in area B for the MRSH are chemically and kinematic consistent with dwarf galaxies.

The overlapping area in the HVTD and MRSH which have same
element abundances means that  they have similar origin. The possible 
origin in these regions were in-situ. 
After undergoing some processes, some stars become halo stars, 
and other stars still maintain the kinematic characteristics of disk stars.
The in-situ population can contain stars formed in the initial gas 
collapse \citep{Samland03}.  \cite{Cooper15} gave two distinct origins of in-situ halo stars:
gas that has been stripped from satellite 
galaxies by tides and ram pressure, and gas is incorporated directly into the smooth halo of the main galaxy
by cosmological infall and Supernova-Driven outflow from the central
galaxy.

In addition, the overlapping area could also include some stars originated from within the Milky Way disk.
\cite{Bonaca17} proposed two disk heating mechanisms to form the metal-rich halo stars: runaway stars 
and radial migration. Runaway stars are young stars that are formed 
in the disk and then ejected from their birthplace.  
Based on the  discussion of metallicity and spectral type in \cite{Bonaca17}, 
combined with the characteristics of our sample,   we can conclude that runaway stars are a minor component of
the observed metal-rich stellar halo.

Radial migration has been recognized as a important component in explaining numerous observations 
such as the spread in the age-metallicity 
relation \citep[e.g.,][]{SB02,Rokar08,Minchev10} and its abundance patterns \citep{SB09a}. 
\cite{El16} proposed star experienced significant radial migration by two 
related processes. First, some stars are formed during
gas outflows, so that their initial orbits can be eccentric and have
large apocenters.  Second, the gravitational potential of the galaxy will have strong fluctuations
under the combination of the inflow gas accretion and gas outflow. 
This fluctuation will affect the stellar orbits, 
which will eventually become heated to a more
isotropic distribution.    Because those stars within the overlapping area
have similar chemical compositions and orbital eccentricities, 
it indicates that they have the same origin and may have 
undergone similar processes.  We consider radial migration is an possible formation mechanism for the stars
in the overlapping area. 

For the stars in the area A, as we shown in Figure~\ref{figure7}, their characteristics are very obvious 
( e.g. high [Fe/H], low eccentricity, low $\alpha$-element ).  
The stars in area A are mainly HVTD, the stellar metallicity in this 
region lies  between -0.2 dex and 0.4 dex. 
These stars in this region have the characteristics of thin disk stars.
According to the proposed formation mechanism of thick disk, 
we infer that these stars could originate from the heating of a pre-existing thin disk.

\begin{figure}[h]
	\includegraphics[width=1\hsize]{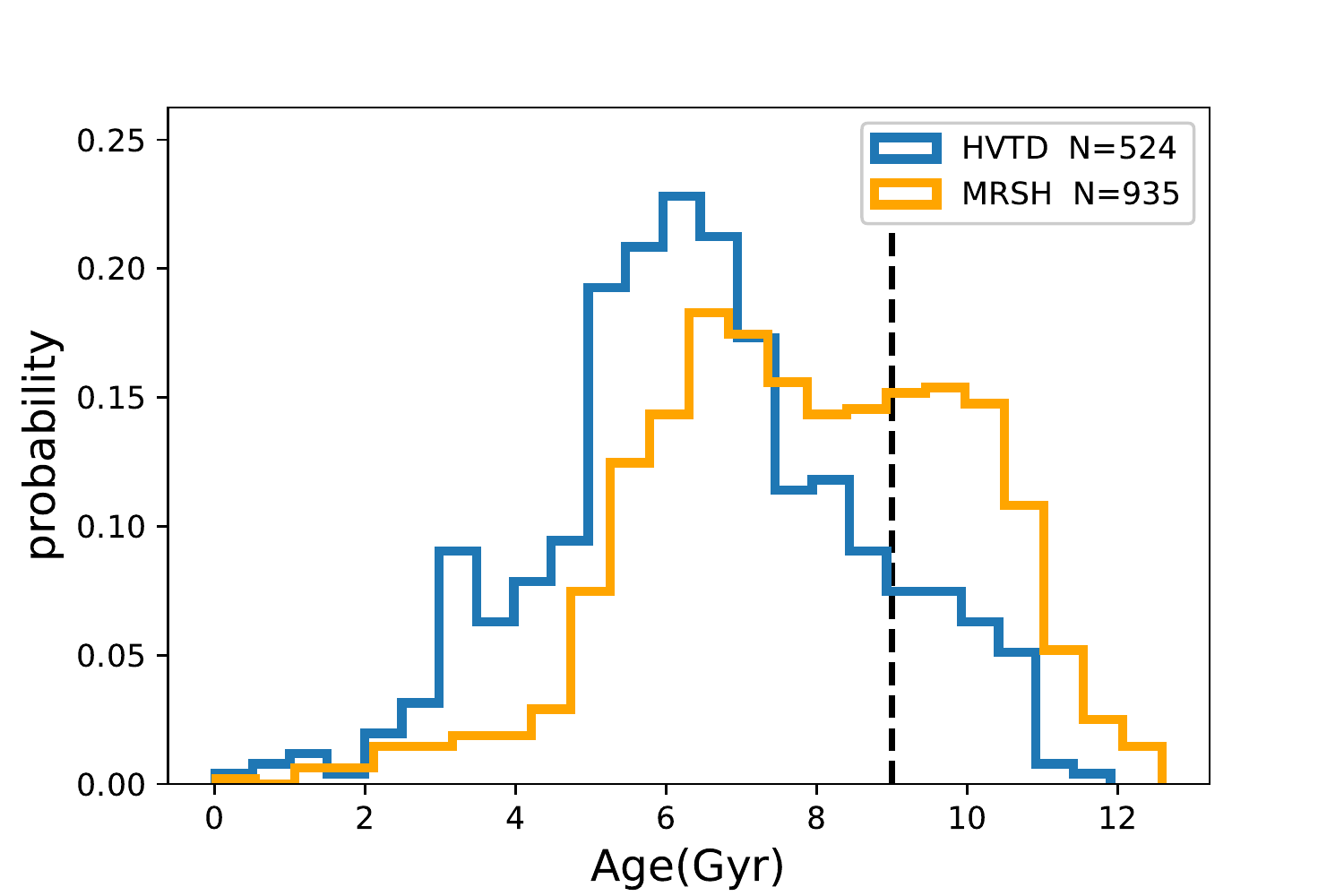}
	\caption{ Age distributions of the MRSH and HVTD, we get through 
	cross-matching our sample and \cite{Sand18}.  We got 524 HVTD samples with blue lines and 935 MRSH samples with orange lines.
	The dash line position is 9 Gyr,	Samples larger than 9 Gyr have experienced the Gaia–Sausage
	merger event as discussion in \cite{Yan20}.}
	\label{figure8}
\end{figure}

The stellar ages also provide important clues to probe the possible origins of the population. 
But it is difficult to obtain the accurate ages of stars. 
Here, we use the age range of the stars to discuss the potential origins of the MRSH and HVTD.
Ages of our sample stars are obtained by cross-matching with catalog of Sanders18 catalog \citep{Sand18}.  
\cite{Sand18} only estimated masses and ages for the stars metal-richer than $-1.5$ dex  and the maximum age isochrone considered is 12.6 Gyr.  
The age distributions in the MRSH and HVTD are shown in Figure \ref{figure8},  which imply that both MRSH and HVTD contain young stars ($<9$ Gyr) and old stars ($>9$ Gyr).
 The young MRSH stars were possible born in-situ, but the Gaia-Sausage merger event may have an important effect on the old stars.  
The HVTD stars also form in an old proto-disk, but these stars may be so less affected by the Gaia-Sausage merger event than MRSH that they remain some properties of the thick disk \citep{Yan20}.

According to above analysis of chemical element abundance and kinematics and age,
we consider that the HVTD and the MRSH each have two origins.  The MRSH stars are formed from the accretion of smaller galaxies and in-situ formation.
The HVTD stars could originate from the heating of a pre-existing thin disk and in-situ formation.

\section{Summary and conclusions}
\label{section5}
Based on the data of APOGEE DR16 and Gaia DR2, we first obtained high-velocity sample stars ($v_{tot}\geq 220 ~{\rm km ~s^{-1}}$). 
According to the distribution of rotational velocity and metallicity ($\rm [Fe/H] >-1$ dex) of this sample stars, we divide this sample stars
into HVTD ($v_{\phi}>90~{\rm km~s^{-1}}$) and MRSH ($v_{\phi}<90~{\rm km~s^{-1}}$) and studied their elements abundance distribution to confirm their origins.

We found the MRSH has relatively low metallicity ($\rm -1.0$ dex, $-0. 4$ dex), 
 but the HVTD not only has a relatively small number in the lower metallicity [Fe/H], 
 but also has more metal-rich stars in metallicity [Fe/H] ($\rm -0. 4$ dex, $0. 4$ dex). 
Furthermore, we use maximum likelihood method to estimate the vertical metallicity gradient of two samples.  
The metallicity vertical gradient of the HVTD is
 $d\text{[Fe/H]}/d|z|=-0. 16\pm 0. 0008 ~{\rm dex/kpc}$ in $z<3$ kpc, and the metallicity 
 vertical gradient of the MRSH is $d\text{[Fe/H]}/d|z|=-0. 02\pm 0. 0003~{\rm dex/kpc}$,  which is a very flat gradient.

The elements abundance of stars can provide important clues to probe their origin. 
In this study,  we selected 11 elements and $\alpha-$element as a function of $\rm [Fe/H]$ 
to study the chemical properties and origins of sample stars. 
In order to comprehensively consider various element abundance indicators, 
we performed dimensionality reduction processing on the data according 
to the t-SNE algorithm.   It is clear that 
the sample is divided into three areas: area A in HVTD, area B in MRSH and overlapping area of HVTD and MRSH, which can relate with some formation mechanisms.
From the chemical element abundance and kinematics and age,
we can conclude that the HVTD and the MRSH each have two origins.  The MRSH stars are formed from the accretion of smaller galaxies and in-situ formation.
The HVTD stars could originate from the heating of a pre-existing thin disk and in-situ formation. 

\section{Acknowledgements}
\par 

We thank especially the referee for insightful comments and suggestions, which has improved the paper significantly. This work was supported by the National Natural Foundation of China (NSFC No. 11973042, No. 11973052 and No. 11873053).  It was also supported by the Fundamental Research Funds for the Central Universities and the National Key R\&D Program of China No. 2019YFA0405501.     H.J.N. acknowledges funding
from US NSF grant AST-1908653. This project was developed in part at the 2016 NYC Gaia Sprint, hosted by the Center for Computational Astrophysics at the Simons Foundation in New York City.
\par 
Funding for the Sloan Digital Sky Survey IV has been provided by the Alfred P. Sloan Foundation, the U.S. Department of Energy Office of Science, and the Participating Institutions. SDSS-IV acknowledges
support and resources from the Center for High-Performance Computing at
the University of Utah. The SDSS web site is www.sdss.org.
\par 
SDSS-IV is managed by the Astrophysical Research Consortium for the
Participating Institutions of the SDSS Collaboration including the
Brazilian Participation Group, the Carnegie Institution for Science,
Carnegie Mellon University, the Chilean Participation Group, the French Participation Group, Harvard-Smithsonian Center for Astrophysics,
Instituto de Astrof\'isica de Canarias, Johns Hopkins University,
Kavli Institute for the Physics and Mathematics of the Universe (IPMU) /
University of Tokyo, Lawrence Berkeley National Laboratory,
Leibniz Institut f\"ur Astrophysik Potsdam (AIP),
Max-Planck-Institut f\"ur Astronomie (MPIA Heidelberg),
Max-Planck-Institut f\"ur Astrophysik (MPA Garching),
Max-Planck-Institut f\"ur Extraterrestrische Physik (MPE),
National Astronomical Observatories of China, New Mexico State University,
New York University, University of Notre Dame,
Observat\'ario Nacional / MCTI, The Ohio State University,
Pennsylvania State University, Shanghai Astronomical Observatory,
United Kingdom Participation Group,
Universidad Nacional Aut\'onoma de M\'exico, University of Arizona,
University of Colorado Boulder, University of Oxford, University of Portsmouth,
University of Utah, University of Virginia, University of Washington, University of Wisconsin,
Vanderbilt University, and Yale University.
\par 
This work has made use of data from the European Space Agency (ESA) mission 
Gaia (http://www.cosmos.esa.int/gaia), processed by the Gaia Data Processing and
Analysis Consortium (DPAC, http://www.cosmos.esa.int/web/gaia/dpac/consortium). 
 Funding for DPAC has been provided by national institutions, 
 in particular the institutions participating in the Gaia Multilateral Agreement.


\begin{thebibliography}{}
	\bibitem[Abadi et al.(2003)]{Ab03}
	Abadi M. G., Navarro J. F., Steinmetz M., \& Eke V. R., 2003, ApJ, 597, 21.


	\bibitem[An et al.(2013)]{An13}
	An D., Beers T. C., Johnson J. A., Pinsonneault M. H., et al., 2013, ApJ,  763,  65

	\bibitem[An et al.(2015)]{An15}
	An D., Beers T. C., Santucci R. M., Carollo D., et al., 2015, ApJL,  813,  L28
	\bibitem[Astraatmadja \& Bailer-Jones(2016a)]{AB16a}
	Astraatmadja T. L., \& Bailer-Jones C. A. L., 2016a, ApJ, 832, 137
	\bibitem[Astraatmadja \& Bailer-Jones(2016b)]{AB16b}
	Astraatmadja T. L., \& Bailer-Jones C. A. L., 2016b, ApJ, 833, 119
	\bibitem[Astropy Collaboration et al.(2018)]{as18}
	Astropy Collaboration, Price-Whelan, A. M., Sipócz, B. M., et al. 2018, AJ,
156, 123
	\bibitem[Astropy Collaboration et al.(2013)]{as13}
	Astropy Collaboration, Robitaille T. P., Tollerud E. J., et al. 2013, A\&A,
558, A33
	\bibitem[Bailer-Jones(2015)]{Bailer-Jones15}
	Bailer-Jones, C. A. L., 2015, PASP, 127, 994
	\bibitem[Bailer-Jones et al.(2018)]{Bailer-Jones et al.(2018)}
	Bailer-Jones C. A. L., Rybizki J. , Fouesneau M., Mantelet G., Andrae R., 2018, AJ, 156, 58
	\bibitem[Beers et al.(2012)]{Beers12}  
	Beers T. C., Carollo D., Ivezi\'c $\check{Z}$., An D., et al., 2012, ApJ,  746,  34

\bibitem[Beers et al.(2002)]{Beers02}  
Beers T. C., Drilling J. S., Rossi S., Chiba M., et al., 2002, AJ,  124, 931

\bibitem[Beers et al.(2014)]{Beers14} 
Beers T. C., Norris J. E., Placco V. M., Lee Y. S., et al., 2014, ApJ,  794,  58

\bibitem[Beers et al.(1995)]{Beers95}  
Beers T. C.,  Sommerlarsen J., 1995, ApJS, 96, 175

	
	
	
	
	
	\bibitem[Belokurov et al.(2018)]{Belokurov18}
	Belokurov V., Erkal D., Evans N. W., Koposov S. E., \& Deason A. J. 2018,
MNRAS, 478, 611
	\bibitem[Belokurov et al.(2019)]{Belokurov19}
	Belokurov V., Sanders J. L., Fattahi A., Smith M. C., et al., 2019
arXiv:1909.04679
       \bibitem[Belokurov et al.(2019)]{Belokurov20}
       Belokurov , V., Sanders, J. L., Fattahi, A., et al., 2020, MNRAS,
doi:10.1093/mnras/staa876

	\bibitem[Bensby et al.(2005)]{Bensby05}  
	Bensby T., Feltzing S., Lundstr{\"o}m I., Ilyin I., 2005, A\&A, 433, 185
	
	\bibitem[Bergemann et al.(2017)]{Bergemann17}
	Bergemann M., Collet R., Sch{\"o}nrich R., Andrae R., et al., 2017, ApJ, 847, 16
	
	\bibitem[Bilir et al.(2012)]{Bilir12}
	Bilir S., Karaali S., Ak S.,  et al., 2012, MNRAS, 421, 3362

	\bibitem[Bird et al.(2020)]{Bird20}
	Bird S. A., Xue X.X., Liu  C., et al., 2020,  arXiv:2005.05980	
		
	\bibitem[Bland-Hawthorn \& Gerhard(2016)]{Bland-Hawthorn16}
	Bland-Hawthorn J., Gerhard O., 2016, ARA\&A, 54, 529

	\bibitem[Bonaca et al.(2017)]{Bonaca17}
	Bonaca A., Conroy C., Wetzel A., Hopkins P. F., et al., 2017, ApJ, 845, 101
	
	\bibitem[Bonaca et al.(2020)]{Bonaca20}
       Bonaca, A., Conroy, C., Cargile, P. A., et al. 2020, arXiv e-prints,
arXiv:2004.11384
	
	\bibitem[Bovy(2015)]{Bovy15}
	Bovy J., 2015, ApJS, 216, 29
	\bibitem[Brook et al.(2004)]{Brook04}
	Brook C. B., Kawata D., Gibson B. K., Freeman K. C., 2004, ApJ, 612, 894
	
	
	\bibitem[Carollo et al.(2007)]{Carollo07}
	Carollo D., Beers T. C., Lee Y. S., Chiba M., et al., 2007, Nature,  450, 1020

	\bibitem[Carollo et al.(2010)]{Carollo10}
	Carollo D., Beers T. C., Chiba M., Norris J. E., et al., 2010, ApJ, 712, 692
	
	\bibitem[Carollo et al.(2012)]{Carollo12}
	Carollo D., Beers T. C.,Bovo J.,et al., 2012, ApJ, 744, 195

	
	\bibitem[Carollo et al.(2019)]{Carollo19}
Carollo D., Chiba M., Ishigaki M., Freeman K., et al., 2019, ApJ,  887, 22	
	
	
	\bibitem[Carollo et al.(2020)]{Carollo20}
	Carollo D., Chiba M., 2020, arXiv 2010.00235v1
	

	\bibitem[Chen et al.(2011)]{Chen11}
	Chen Y. Q., Zhao G., Carrell K., \& Zhao J. K. 2011, ApJ, 142, 184
	\bibitem[Chiba \& Beers (2000)]{Chiba00}
	Chiba M., Beers T. C., 2000, AJ, 119, 2843
	\bibitem[Cooper et al.(2015)]{Cooper15}
	Cooper A. P., Parry O. H., Lowing B., Cole S., et al., 2015, MNRAS, 454,
3185
        \bibitem[Conroy et al.(2019)]{Conroy19}
Conroy C., Naidu R. P., Zaritsky  D., et al., 2019, ApJ, 887, 237

\bibitem[Deason et al.(2018)]{Deason18}
Deason A. J., Belokurov V., Koposov S. E., Lancaster L., 2018, ApJL,  862,  L1
\bibitem[Eisenstein et al.(2011)]{Eisenstein2011}
Eisenstein, D. J., Weinberg, D. H., Agol, E., et al. 2011, AJ, 142, 72
\bibitem[El-Badry et al.(2016)]{El16}
El-Badry K., Wetzel  A., Geha M., et al. 2016, ApJ, 820, 131
	\bibitem[Fattahi et al.(2019)]{Fattahi19}
	Fattahi  A., Belokurov V., Deason  A. J., et al. 2019, MNRAS, 484, 4471
	\bibitem[Feuillet et al.(2019)]{Feuillet2019}
	Feuillet, D. K., Frankel, N., Lind, K., et al. 2019, MNRAS, 489, 1742
	\bibitem[Foreman-Mackey et al.(2013)]{for13}
	Foreman-Mackey, D., Hogg, D. W., Lang, D., \& Goodman, J. 2013, PASP,
125, 306
\bibitem[Gaia Collaboration et al.(2018a)]{Gaia18a}
Gaia Collaboration, Brown A. G. A., Vallenari A., et al., 2018a, A\&A, 616, A1

\bibitem[Gaia Collaboration et al.(2018b)]{Gaia18b}
Gaia Collaboration, Katz D., Antoja T., et al., 2018b, A\&A, 616, A11	
	
	
	\bibitem[Gallart et al.(2019)]{gall19}
	Gallart C., Bernard E. J., Brook C. B., Ruiz-Lara T., et al., 2019
arXiv:1901.02900
          

      \bibitem[Gandhi \& Ness (2019)]{Gandhi19}
     Gandhi, S. S., Ness, M. K., 2019, ApJ, 880,134
	\bibitem[Grand et al.(2020)]{Grand20}
	 Grand, R. J. J., Kawata, D., Belokurov, V., et al. 2020, MNRAS, 497, 1603
	\bibitem[Gu et al.(2015)]{Gu15}
	Gu J. Y., Du C. H., Jia Y. P., Peng X. Y., et al., 2015, MNRAS,  452, 3092
	
	\bibitem[Gu et al.(2016)]{Gu16}
	Gu J. Y., Du C. H., Jing Y. J., Zuo W. B., 2016, ApJ,  826,  36
	
	\bibitem[Gu et al.(2019)]{Gu19}
	Gu J. Y., Du C. H., Zuo W. B., 2019, ApJ,  877,  83
	
	\bibitem[Han et al.(2020)]{Han20}
	Han D. R., Lee Y. S., Kim Y. K., Beers T. C., 2020, ApJ, 896, 14
	\bibitem[Hawkins et al.(2015)]{Hawkins15}
	Hawkins K., Kordopatis G., Gilmore  G., et al. 2015, MNRAS, 447, 2046
	\bibitem[Haywood et al.(2018)]{Hayw18}
	Haywood M., Di Matteo P., Lehnert M. D., Snaith O., et al., 2018, ApJ, 863,
113
	\bibitem[Hayes et al.(2018)]{Hayes18}
	Hayes C. R., Majewski S. R., Hasselquist S., et al., 2018, ApJ, 859, 8
	
	\bibitem[Helmi et al.(2018)]{Helmi18}
Helmi A., Babusiaux C., Koppelman H. H., Massari D., et al., 2018, Nature,  563, 85
	
	\bibitem[Ivezi\'c et al.(2012)]{Ivezi12} Ivezi\'c  \v{Z}., Beers T. C., Juri\'{c}  M.,2012, ARA\&A, 50, 251
	
	\bibitem[Jing et al.(2016)]{Jing16}
       Jing Y. J., Du C. H., Gu J. Y., Jia Y. P., et al., 2016, MNRAS, 463, 3390
		
	\bibitem[J{\"o}nsson et al.(2020)]{Jonsson20}
	J{\"o}nsson, H., Holtzman, J. A., Allende Prieto, C., et al., 2020, AJ, submitted
	
	\bibitem[Kinman et al.(2012)]{Kinman12}
	Kinman T. D., Cacciari C., Bragaglia A., Smart R., et al., 2012, MNRAS,  422, 2116
	
\bibitem[Koppelman et al.(2019)]{Koppelman19}	
     Koppelman  H. H.,  Helmi A., Massari D., et al., 2019, A\&A, 631, 9	
\bibitem[Kordopatis et al.(2020)]{Kordopatis20}
	 Kordopatis G., Recio-Blanco A., Schultheis M., et al., 2020,  A\&A, 643, A69
\bibitem[Kruit \& Freeman(2011)]{Kruit11} van der Kruit P. C., Freeman K.C., 2011, ARA\&A, 49, 301


\bibitem[Lancaster et al.(2019)]{Lancaster19}
Lancaster L., Koposov S. E., Belokurov V., Evans N. W., et al., 2019, MNRAS,  486, 378

	
	\bibitem[Letarte et al.(2010)]{Letarte10}
	Letarte  B., Hill V., Tolstoy E., et al., 2010, A\&A, 523, 17
	\bibitem[Li \& Zhao (2017)]{Li17}
	Li C. D., Zhao G., 2017, ApJ, 850, 25
	\bibitem[Liu \& van de Ven.(2012)]{Liu12}
	Liu C., \& van de Ven, G., 2012,  MNRAS, 425(3), 2144.
	\bibitem[Liu et al.(2018)]{Liu18}
	Liu S., Du C. H., Newberg H. J., Chen Y. Q., et al., 2018, ApJ,  862, 163
	\bibitem[Luri et al.(2018)]{luri18}
	Luri X., Brown  A. G. A., Sarro  L. M., et al., 2018, A\&A, 616, A9
	\bibitem[Martig et al.(2016)]{Martig16}
	Martig M., Fouesneau M., Rix H.-W., et al., 2016, MNRAS, 456, 3655
	\bibitem[ Masseron \& Gilmore (2015)]{Masseron15}
	Masseron T., \& Gilmore G. ,2015, MNRAS, 453(2), 1855.
	
	
	\bibitem[McMillan(2017)]{McMillan17}
	McMillan P. J., 2017, MNRAS, 465, 76
	
	
\bibitem[Myeong et al.(2018)]{Myeong18}
Myeong G. C., Evans N. W., Belokurov V., Sanders J. L., et al., 2018, ApJL,  863,  L28

\bibitem[Myeong et al.(2019)]{Myeong19}
Myeong G. C., Vasiliev E., Iorio G., Evans N. W., et al., 2019, MNRAS,  488, 1235
	
	
	\bibitem[Mikolaitis et al.(2014)]{Mikolaitis14}
	Mikolaitis \v{S}, Hill V., Recio-Blanco A., et al., 2014, A\&A, 572, A33
	\bibitem[Minchev \& Famaey (2010)]{Minchev10}
	Minchev I., \& Famaey B. 2010, ApJ, 722, 112
	
	\bibitem[Morrison et al.(1990)]{Morrison90}
      Morrison H. L., Flynn C., Freeman K. C., 1990, AJ, 100, 1191	
	
	\bibitem[Naidu et al.(2020)]{Naidu20}
	Naidu R. P., Conroy C., Bonaca  A., et al., 2020, ApJ, 901,48
	
	\bibitem[Nissen \& Schuster(2010)]{Nissen10}
	Nissen P. E., Schuster W. J., 2010, A\&A, 511, L10
	\bibitem[Nissen \& Schuster(2011)]{Nissen11}
	Nissen P. E., Schuster W. J., 2011, A\&A,  530,  A15
	\bibitem[Pedregosa et al.(2011)]{Pedregosa11}
	Pedregosa, F., Varoquaux, G., Gramfort, A., et al. 2011, J. Mach. Learn. Res.,
12, 2825
	\bibitem[Posti et al.(2018)]{Posti18}
	Posti L., Helmi A., Veljanoski J., Breddels M. A., 2018, A\&A,  615,  A70
	\bibitem[Peng et al.(2012)]{Peng12}
	Peng X.Y., Du C.H., Wu Z.Y., 2012, MNRAS, 422, 2756
	
	\bibitem[Peng et al.(2013)]{Peng13}
	Peng X.Y., Du C.H., Wu Z.Y., et al., 2013, MNRAS, 434, 3165
	
	\bibitem[Peng et al.(2018)]{Peng18}
	Peng X.Y., Wu Z.Y., Qi Z.X., et al., 2018, PASP, 130, 4102

	\bibitem[Prochaska et al.(2000)]{Prochaska00}
	Prochaska J. X., Naumov S. O., Carney B. W., McWilliam A., Wolfe A. M., 2000, AJ, 120, 2513
	\bibitem[Quinn et al.(1993)]{Quinn93}
	Quinn, P. J., Hernquist, L., \& Fullagar, D. P. 1993, ApJ, 403, 74
	\bibitem[Ro{\v s}kar et al.(2008)]{Rokar08}
	Ro{\v s}kar R., Debattista V. P., Quinn T. R., Stinson G. S., \& Wadsley J. 2008,
ApJL, 684, L79
       \bibitem[\c{S}ahin \& Bilir (2020)]{Sahin20}
       \c{S}ahin T.,  Bilir S., 2020, ApJ, 899,411
       
       
	\bibitem[Sales et al.(2009)]{Sales09}
         Sales L. V., Helmi  A., Abadi M. G., et al. 2009, MNRAS, 400, 61
	\bibitem[Samland \& Gerhard (2003)]{Samland03}
	Samland M., Gerhard O. E., 2003, A\&A, 399, 961
	\bibitem[Sanders \&Das(2018)]{Sand18}
	Sanders J. L., Das P., 2018, MNRAS, 481, 4093
	
	\bibitem[Sellwood \& Binney (2002)]{SB02}
	Sellwood J. A., \& Binney J. J., 2002, MNRAS, 336, 785
	
	\bibitem[Sch{\"o}enrich \& Binney (2009a)]{SB09a}
Sch{\"o}enrich R., \& Binney J. J., 2009, MNRAS, 396, 203

\bibitem[Sch{\"o}enrich \& Binney (2009b)]{SB09b}
Sch{\"o}enrich R., \& Binney J.J., 2009, MNRAS, 399,1145

\bibitem[Sch{\"o}enrich et al.(2011)]{Sch11}	
Sch{\"o}enrich R., Asplund M., Casagrande L., 2011, MNRAS 415, 3807
\bibitem[Steinmetz et al.(2006)]{stei06}
Steinmetz M., Zwitter T., Siebert A., et al. 2006, AJ, 132, 1645

	
	
	\bibitem[Tun\c{c}el G{\"u}\c{c}tekin et al.(2019)]{Tun19}
	Tun\c{c}el G{\"u}\c{c}tekin, S., Bilir, S., Karaali, S., et al., 2019, AdSpR, 63, 1360
	\bibitem[Van der Maaten \& Hinton (2008)]{van08}
	Van der Maaten, L., \& Hinton, G., 2008,  The Journal of Machine Learning Research, 9, 85
	
	\bibitem[Villalobos \& Helmi(2008)]{Villalobos08}
	Villalobos $\rm \acute{A}$., Helmi A., 2008, MNRAS, 391(4), 1806
	
	\bibitem[Wu et al.(2021)]{Wu21}
	Wu Y. Q., Xiang M.S., Chen Y.Q., et al., 2021, MNRAS, 501, 4917
	
	\bibitem[Xiang et al.(2015)]{Xiang15}
	Xiang M.S., Liu X.W., Yuan H.B., et al., 2015, Res. Astron. Astrophys., 15, 1209
	
	\bibitem[Xiang et al.(2017)]{Xiang17}
	Xiang M.S., Liu X.W., Shi J.R., et al., 2017, ApJS, 232, 2
	
	\bibitem[Yan et al.(2020)]{Yan20}
	Yan Y. P., Du C. H., et al., 2020, ApJ, 903, 131
	\bibitem[Yan et al.(2019)]{Yan19}
	Yan Y. P., Du C. H., et al., 2019, ApJ, 880, 36
	\bibitem[Yuan et al.(2020)]{Yuan20}
	Yuan Z., Myeong G.C., Beers T., et al., 2020, ApJ, 891, 39
	
	\bibitem[Zasowski et al.(2013)]{Zasowski13}
        Zasowski G., Johnson J. A., Frinchaboy P. M., Majewski S. R., et al., 2013, AJ,  146,  81

       \bibitem[Zhao et al.(2012)]{zhao12}
       Zhao G., Zhao Y. H., Chu Y. Q., Jing Y. P., et al. 2012, RAA, 12, 723
	\bibitem[Zuo et al.(2017)]{Zuo17}
	Zuo W., Du C. H., Jing Y. J., Gu J. Y., et al., 2017, ApJ,  841, 59
	
	\end{thebibliography}
\end{document}